\documentclass[aip,reprint,superscriptaddress]{revtex4-1}
\usepackage{graphicx}
\usepackage{dcolumn}
\usepackage{bm}
\usepackage[utf8]{inputenc}
\usepackage[T1]{fontenc}
\usepackage{mathptmx,amsmath,amssymb}
\usepackage{xcolor}

\begin{document}
\providecommand{\U}[1]{\protect\rule{.1in}{.1in}}

\title{Observation of Planar Hall Effect in Topological Insulator -- Bi$_2$Te$_3$}
\author{Archit Bhardwaj}
 \affiliation{Tata Institute of Fundamental Research, Hyderabad, Telangana 500046, India}
 \author{Syam Prasad P.}
 \affiliation{Indian Institute of Technology Hyderabad, Sangareddy, Telangana  502285, India}
\author{Karthik V. Raman}
\affiliation{Tata Institute of Fundamental Research, Hyderabad, Telangana 500046, India}
\author{Dhavala Suri$^1$}
\altaffiliation[Authors to whom correspondence should be addressed:] {kvraman@tifrh.res.in, dhavalas@tifrh.res.in}

\begin{abstract} 

Planar Hall effect (PHE) in topological insulators (TIs) is discussed as an effect that stems mostly from conduction due to topologically protected surface states. Although surface states play a critical role and are of utmost importance in TIs, our present study in Bi$_2$Te$_3$ thin films reflects the need for considering the bulk conduction in understanding the origin of PHE in TIs. This necessity emerges from our observation of an unconventional increase in PHE signal with  TI thickness and temperature where the bulk effect takes over. Here we find an enhancement in PHE amplitude by doubling the Bi$_2$Te$_3$ film-thickness on Si (111) substrate-- from $\approx$~1.9 n$\Omega$m in 14 quintuple layer (QL) to $\approx$~3.1 n$\Omega$m  in 30 QL devices at B = 5~T. Also, the PHE amplitude in the 30 QL Bi$_2$Te$_3$ films grown on two different substrates \textit{viz.} Si (111) and Al$_2$O$_3$ (0001) show an  increase with temperature. Our experiments indicate that the contribution of bulk states to PHE in TIs could be significant. 
\end{abstract}

\maketitle
Topological insulators (TIs) have been the forerunners of research in condensed matter physics due to the plethora of possibilities that surface states offer for fundamental exploration \cite{ando,RevModPhys.83.1057,RevModPhys.82.3045}. However, it is worthwhile to investigate the mesoscopic transport dominated by the bulk states \cite{PhysRevB.96.125125,PhysRevB.98.165408,Barreto2014,PhysRevLett.106.196801}. A phenomenon that has evoked  widespread interest in recent times is the planar Hall effect (PHE), which is development of  transverse voltage in response to  longitudinal current, under  external in-plane magnetic field, in a configuration where transverse voltage due to Lorentz force is zero. The discovery of  PHE dates back to 1954 \cite{PhysRev.94.1121},  and has been used extensively  for its application to Hall sensors \cite{epshtein2002}. PHE based magnetic random access memory devices are also known for  memory storage applications \cite{bason}. Hence,  non-magnetic materials exhibiting PHE are potential candidates for such applications since they eliminate spurious fringe field effects. PHE has been observed in a variety of systems  such as ferromagnet/normal metal bilayers \cite{doi:10.1063/1.4945324,Safranski2019}, ferromagnetic semiconductors \cite{LEE200914,PhysRevB.100.134441}, ferromagnetic metals \cite{EPSHTEIN200380} and topological superconductors \cite{doi:10.1063/1.5063689,10.1093/nsr/nwaa163}. Recently, the effect is in the spotlight due to its role  in probing topological characteristics  such as chirality.  In Weyl semimetals, the origin of PHE coupled with negative magnetoresistance is considered  a signature of chiral anomaly \cite{PhysRevB.96.041110,PhysRevB.83.205101,Li2016,Xiong413,Hirschberger2016, PhysRevB.100.205128,PhysRevLett.119.176804}.
The  origin of  non-zero off-diagonal  terms in the resistivity tensor could be attributed to several mechanisms including classical orbital magnetoresistance \cite{doi:10.1063/1.5133809}. However,  conventional PHE refers to transverse and longitudinal resistivity under in-plane magnetic field, given by eqn (\ref{phe_eqn}) below, where periodicity of PHE is $\pi$:
  \begin{align}
  \begin{split}
    \rho_{xy} &= (\rho_{\parallel} - \rho_{\perp}) \sin \phi \cos \phi,
    \\
    \rho_{xx} &= \rho_{\perp} + (\rho_{\parallel} - \rho_{\perp}) \cos^2\phi
  \end{split}
  \label{phe_eqn}
  \end{align}
where $\phi$ is the angle between current and magnetic field direction lying in the same plane, $\rho_{xy}$ and $\rho_{xx}$ are transverse and longitudinal resistivity respectively. $\rho_{\parallel}$ and $\rho_{\perp}$ are longitudinal resistivities when  $\phi$ is 0$^{\circ}$ and 90$^{\circ}$ respectively.

PHE in topological insulators is believed to stem from surface state conduction. This was demonstrated by  Taskin et. al. \cite{Taskin2017}  where the PHE amplitude measured across the Fermi level exhibited local maxima on either side of the Dirac point in the surface state regime. Bulk crystals of Sn doped Bi$_{1.1}$Sb$_{0.9}$Te$_2$S   \cite{doi:10.1063/1.5031906} also exhibit an oscillating PHE which appears only in the  topologically protected surface state regime. It is interesting to note that there is no experimental report of conventional PHE in Bi$_2$Se$_3$ despite  being   extensively studied \cite{PhysRevB.94.081302,Xia2009,Analytis2010}. However, it does exhibit a non-linear unconventional PHE \cite{PhysRevLett.123.016801}, which refers to PHE with periodicity of 2$\pi$.  Such an unconventional PHE is also observed by Rakhmilevich et. al., \cite{PhysRevB.98.094404} in   Bi$_{0.22}$Sb$_{0.78}$Te$_3$/EuS, where induced ferromagnetism shows anisotropic magnetoresistance along with PHE of periodicity 2$\pi$.   Following these experiments several groups have worked on explaining the results theoretically: (i)~S-H Zheng et. al. \cite{PhysRevB.101.041408} show that PHE can arise as a consequence of anisotropic backscattering from the Dirac cone, tilted due to in-plane magnetic field. Including non-linear terms in momentum to the surface state Hamiltonian explains the experimental result by Taskin et. al., without the need to invoke scattering by impurities. (ii) Further, S. Nandy et. al. \cite{Nandy2018} model PHE in TIs in the bulk conduction limit using the Boltzmann transport equation, and notice that  PHE arises purely from the Berry curvature of  bulk bands. (iii) A recent calculation \cite{soori2021finite} using scattering theory shows  that PHE in surface states of a TI  stems from the transverse displacement of the dispersions under in-plane magnetic field and extends the calculations to spin orbit coupled systems as well \cite{soori2021finite2}. We recognize that a comprehensive satisfactory understanding of PHE in topological insulators is yet to be developed. We attempt to address this and present our experimental results.  

\par In this article, we performed PHE experiments in thin films of Bi$_2$Te$_3$ by rotating the magnetic field in-plane by an angle $\phi$ with respect to the direction of current. Bi$_2$Te$_3$\cite{yao,Zhang2009,PhysRevB.90.075105, PhysRevLett.103.266801,CONCEPCION201961,KRUMRAIN2011115,doi:10.1063/1.4815972,Analytis2010,Qu821} is a three-dimensional (3-D) TI with a small bulk band gap $<$~0.15eV  \cite{doi:10.1063/1.4955188} owing to  which it is experimentally challenging to separate the bulk and surface contributions to the transport.  An in-plane magnetic field has no orbital effect on the surface of the TI, but does cause a transverse shift in the Dirac cone. Hence, the   development of a response to in-plane magnetic field is intriguing \cite{PhysRevB.83.245428,PhysRevB.86.165404}. It is an interesting question to ask if PHE can be observed in a bulk conduction dominated 3D-TI. To answer this question, we examined the transport dynamics of Bi$_2$Te$_3$ whose relatively small bulk band gap allows a significant mixing of the bulk and surface state signals. We investigate PHE in Bi$_2$Te$_3$ films of two different thicknesses grown on different substrates.  The following sections describe the growth of the films, the results of measurements and discuss the plausible mechanism.

 High quality epitaxial films of Bi$_2$Te$_3$ were deposited on Si (111) and Al$_2$O$_3$ (0001) substrates using molecular beam epitaxy technique at a base pressure $\approx$ 10$^{-9}$ mTorr. A constant rate ratio of 1:10 was maintained during growth of all the samples, to ensure Te rich growth atmosphere. A two-step growth process was employed. Substrate temperature was maintained at 230~$^{\circ}$C for Al$_2$O$_3$ (0001) and 200 $^{\circ}$~C for Si (111)   respectively, optimized to attain epitaxial film for the first 4 QL. The samples were then annealed at 270$^{\circ}$~C for 30 min and the remaining thickness of the sample was grown at 270$^{\circ}$~C. Thicknesses of 14 quintuple layers (QL) and 30 QL were grown on both substrates. All the samples were capped with 2~nm of Te followed by 5~nm of  Al$_2$O$_3$. Transport measurements were performed in a variable temperature insert cryostat coupled  with Attocube rotation stage for angle dependent scans; samples could reach a stable lowest temperature of 1.5~K. Keithley current source (6221) and nanovoltmeter (2182A) were used for DC resistance measurements. All the samples were manually patterned into Hall bars for measurements, thus avoiding any contamination due to lithographic processes. For rest of  the manuscript we use the nomenclature of the samples: BSi14 and BSi30  for Bi$_2$Te$_3$ on Si~(111) of 14~QL and 30 QL respectively, BA14 and BA30 for Bi$_2$Te$_3$ on Al$_2$O$_3$ (0001)  of 14~QL and 30 QL respectively.

\begin{figure}[tbh]
    \centering
    \includegraphics[width=8cm]{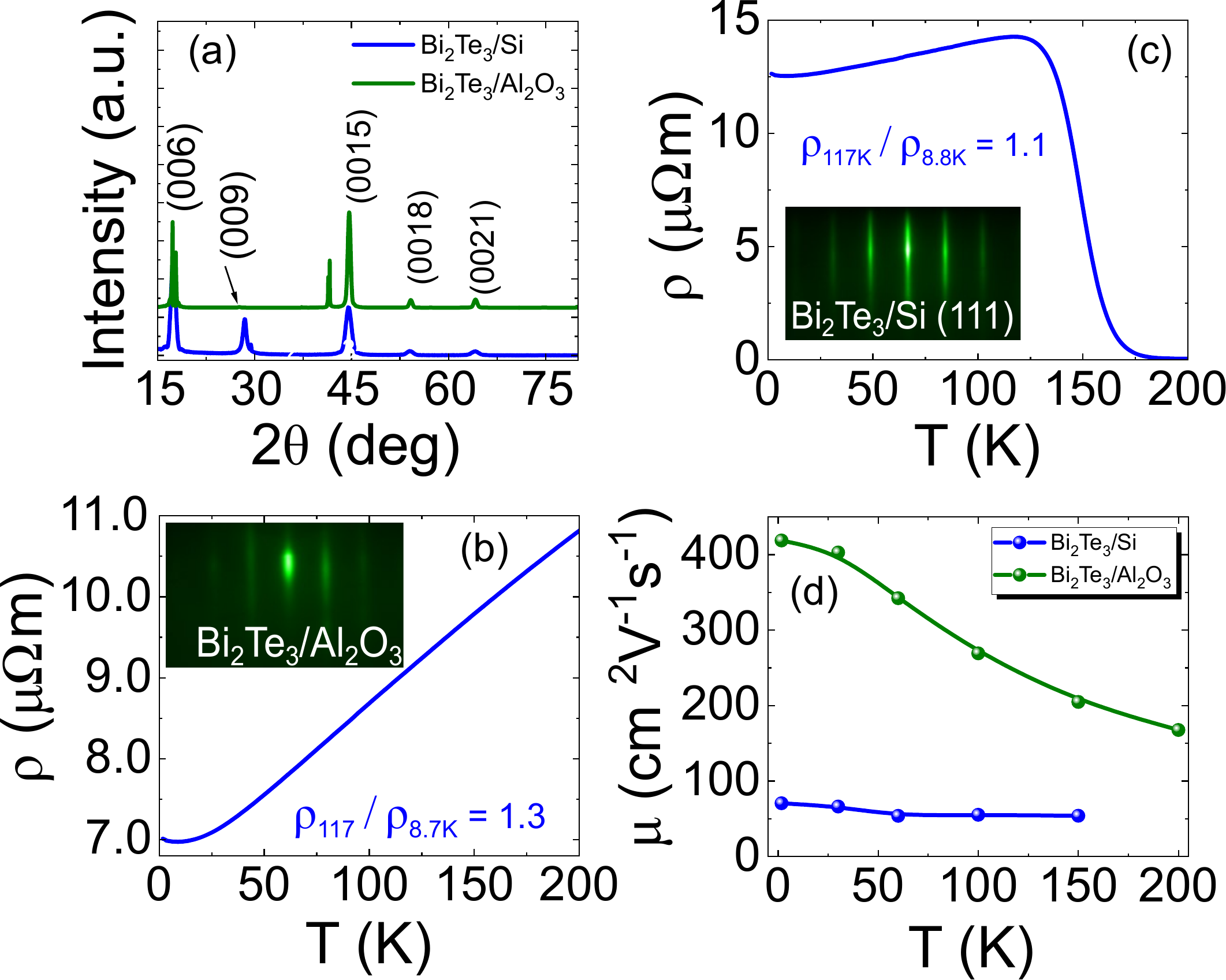}
    \caption{(a) XRD spectrum of Bi$_2$Te$_3$ grown on Si (111) and Al$_2$O$_3$ (0001). Resistivity versus temperature and RHEED images (inset) of 14 QL Bi$_2$Te$_3$ grown on  (b) Al$_2$O$_3$ (0001)  (c) Si (111) (d) Mobility versus temperature for BSi14 (blue) and BA14 (green).}
    \label{basic}
\end{figure}

 Basic characteristics of the films were analysed by  XRD spectra (fig.\ref{basic} (a)) which show sharp characteristic Bi$_2$Te$_3$  peaks. Raman spectra of the films show characteristic Bi$_2$Te$_3$ phase as well (refer to supplementary info). The resistivity versus temperature curves (fig. 1 (b)-(c)) of the two films show metallic nature. The minor upturn in resistance at low temperatures for T $<$ 10~K may be attributed to significant  contribution from impurity bands   \cite{PhysRevB.86.045314,PhysRevB.82.241306} or the e-e interaction.  RHEED images shown in the inset exhibit highly epitaxial growth. We also remark that both films show hexagonal symmetry corresponding to the substrate indicative of epitaxial growth. Mobility calculated from the Hall measurements for BA14 is higher than that of BSi14~(fig.~\ref{basic} (d)).   This maybe attributed to compressive strain on the films due to lattice mismatch which is 9\% for Bi$_2$Te$_3$ grown on  Al$_2$O$_3$ and 18\% on Si substrates \cite{PhysRevB.84.085106, BRAHLEK201554}.
\begin{figure} [tbh]
\centering
    \includegraphics[width=8cm]{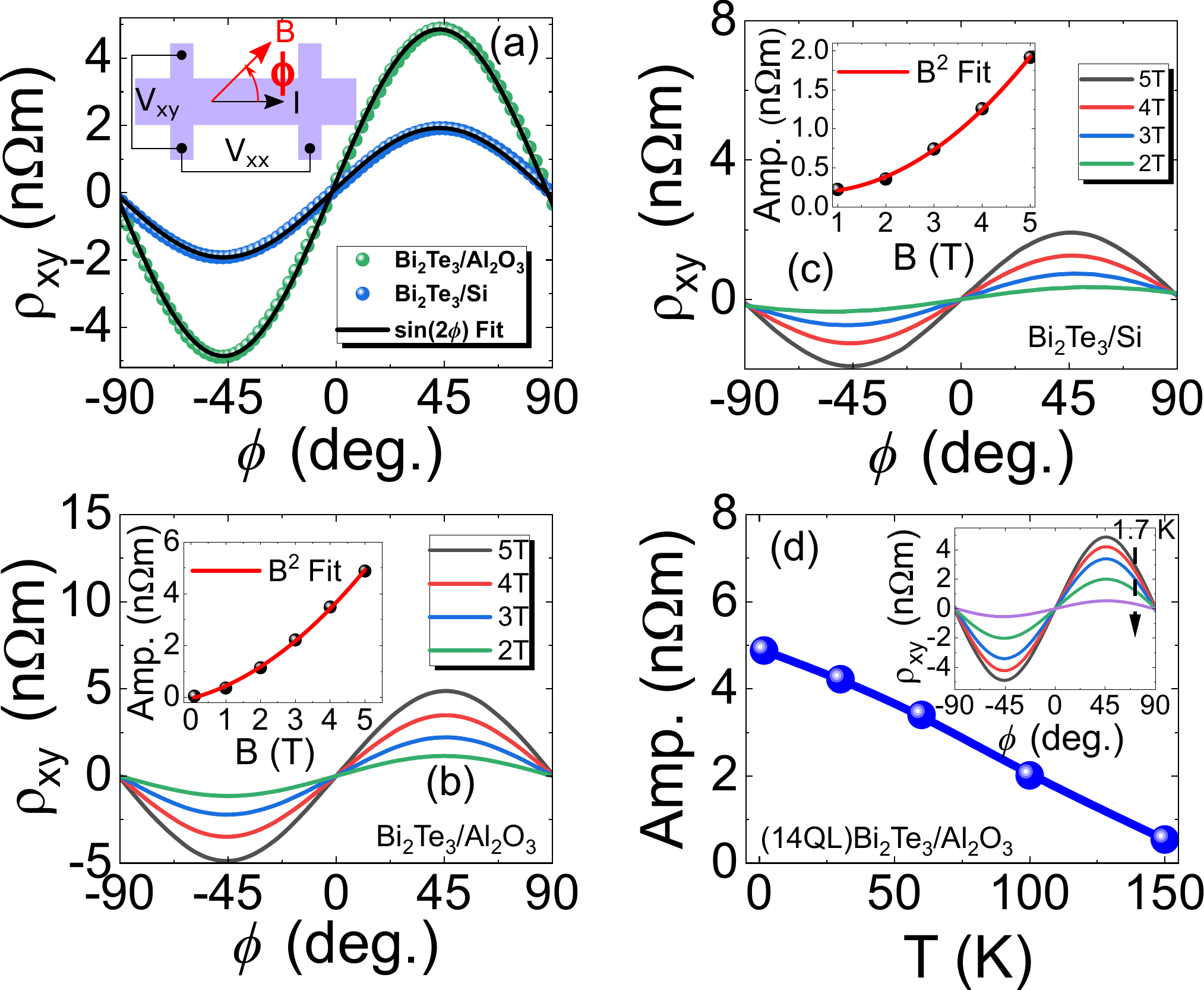}
    \caption{(a) $\rho_{xy}$ vs azimuthal angle ($\phi$)  for BSi14 (blue) and BA14 (green) with $\sin 2\phi$ fit (black line) at B = 5~T, inset shows the schematic of measurement geometry. $\rho_{xy}$ vs $\phi$ at different magnetic field for (b) BA14 (c)~BSi14, insets show B$^2$ fit. (d)  PHE amplitude ($\rho_{xy} (\phi = 45^{\circ})$)  for BA14 at B = 5~T. Inset shows the corresponding $\rho_{xy}$ vs $\phi$~.}
    \label{PHE}
\end{figure}    \\

The key result of our work is the demonstration of  planar Hall effect in Bi$_2$Te$_3$ and its uncommon response to temperature and film-thickness. We measure the transverse resistivity as a function of $\phi$ (inset of fig.\ref{PHE} (a)). $\rho_{xy}$ varies as $\sin 2\phi$ with a period of  $\pi$ (fig.\ref{PHE} (a))), as expected for the case of conventional PHE. We observe that PHE amplitude ($\rho_{xy}$ at $\phi = 45$) varies monotonically with field and fits to $B^2$ functional form (inset of fig.\ref{PHE} (b-c)). The longitudinal magnetoresistance ($\rho_{xx}$) varies as $\cos^2 \phi$ consistent with eqn.~(\ref{phe_eqn}) for conventional PHE (refer to suppl. info). The PHE amplitude monotonically decreases with increase in temperature (fig.~\ref{PHE}~(d)) indicating that either (i) surface state contribution to PHE signal is decreasing with temperature or (ii) bulk disorder-induced thermal excitations decrease the PHE signal from the bulk states of BA14 and BSi14 devices (refer to supp. info PHE amplitude vs T of Bi$_2$Te$_3$/Si~(111)). In the low mobility BSi14, either or both of the above effects may dominate causing a decrease in PHE amplitude with temperature. However, in the case of BA14 devices having 6 times higher mobility than in BSi14, the contribution of the latter effect may be less, suggesting a more dominant surface contribution to the total PHE amplitude. This may also explain the larger magnitude of PHE amplitude in BA14 compared to BSi14. Experiments by Taskin et. al. \cite{Taskin2017} and B. Wu et. al. \cite{doi:10.1063/1.5031906} have demonstrated that PHE dominated by surface states decay with temperature  in Bi$_{2-x}$Sb$_x$Te$_3$ and Bi$_{1.1}$Sb$_{0.9}$Te$_2$S, respectively.  Alternatively, the difference in PHE amplitude of the two devices can be explained from the chemical potential arguments. The chemical potential in BA14 is lower than BSi14 \cite{Taskin2017} suggesting a larger surface contribution; calculated using free-electron approximation (m$^* =$0.1 m$_e$, where m$_e = 9.1 \times 10^{-31}$~kg and m$^*$ is the effective mass) to be lying 0.34 eV and 0.58 eV above the conduction band-edge for BA14 and BSi14 devices, respectively. We remark that this is only a crude approximation since it does not consider the precise dispersion relation for each sample.

\begin{figure} [tbh]
    \centering
    \includegraphics[width=8.5cm]{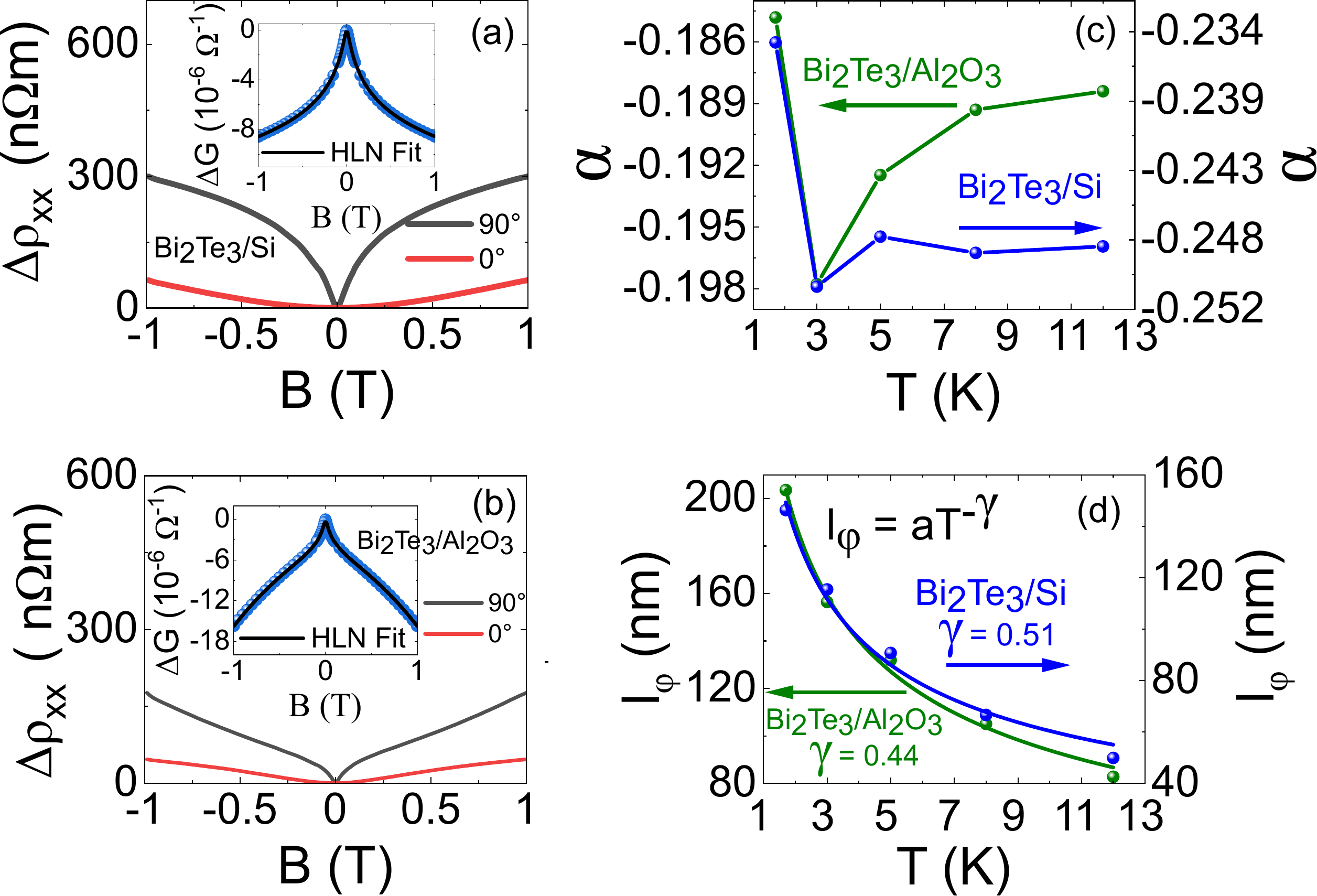}
    \caption{$\rho_{xx}$ vs B  at out-of-plane and in-plane magnetic field, for (a) BSi14 (b) BA14. Insets show magneto-conductance with HLN fit.  HLN parameters extracted from out of plane magneto-transport measurement (inset) for both devices (c) Pre-factor ($\alpha$) vs temperature (d) Coherence length (l$_{\varphi}$) versus temperature. $\Delta\rho_{xx}$~=~$\rho(B) - \rho(0)$.}
    \label{WAL}
\end{figure}

It is fruitful to study the magneto-transport in the out-of-plane magnetic field direction,  to understand the interplay of scattering from scalar impurities and spin-orbit coupling \cite{ma10070807}. As shown in fig.~\ref{WAL} (a) and (b), the resistivity of the sample shows a cusp at low field when the sample is perpendicular to the plane of magnetic field, indicating weak anti-localization (WAL). By rotating the sample from out-of-plane to in-plane magnetic field configuration ($\theta$ is rotated from 90$^{\circ}$ to 0$^{\circ}$), we find that the cusp vanishes and resistivity decreases.  The  magneto-conductance cusp  (insets of fig.~\ref{WAL}) (a)-(b) is fit to Hikami-Larkin-Nagaosa model (eqn.~\ref{HLN}) \cite{10.1143/PTP.63.707} that best explains the scattering and quantifies the coherence length (l$_{\varphi}$) and $\alpha$  whose values indicate the number of conducting 2D channels in the transport and the mechanism responsible for dephasing respectively. 

\begin{equation}
\Delta G = \sigma(H)-\sigma(0) = -\alpha \frac{e^2}{\pi h} \left\{\Psi\left(\frac{1}{2}+\frac{B_\phi}{H}\right)-\log\left(\frac{B_\phi}{H}\right)\right\}
\label{HLN}
\end{equation}
where $B_{\phi} = \frac{h}{8\pi e \text{l}_{\varphi}^2}$ and $\Psi$ is the digamma function \footnote{refer to suppl. info. for details on fitting routine}.  The coherence length  l$_{\varphi}$ at 1.7~K is $\approx$ 160 nm and 200 nm for BSi14 and BA14 respectively (fig. \ref{WAL} (d)).  The power law l$_{\varphi} = aT^{\gamma}$ fit to the temperature dependence of l$_{\varphi}$ shows that the dephasing mechanism is likely to be e-e interaction  \cite{Liu2019}. $\alpha$ (fig.~\ref{WAL} (c)) is $\approx$ 0.19  and 0.23 for BA14 and BSi14, respectively. In the ideal case  the value of $\alpha$ is expected to be 1/2, for conduction due to single coherent channel. However, the observed lower values of $\alpha$ indicate that the bulk conduction dominates the transport in our Bi$_2$Te$_3$ samples irrespective of substrate.  WAL signal decays for $T > 10 K$, however the PHE persists up to much higher temperatures, implying the origin of the two effects are uncorrelated.  For a comprehensive picture of the various parameters of the  Bi$_2$Te$_3$ devices discussed in the manuscript, we illustrate the comparison in table \ref{table}.

\setlength{\arrayrulewidth}{0.25mm}
\setlength{\tabcolsep}{1pt}
\renewcommand{\arraystretch}{1}
\begin{table}[tbh]
\centering
\begin{tabular}{ccccc}
\hline
\hline
Thickness & \multicolumn{2}{c}{14QL} & \multicolumn{2}{c}{30QL}\\
Substrate & Al$_2$O$_3$ & Si & Al$_2$O$_3$ & Si\\
\hline
Mobility (cm$^2$V$^{-1}$s$^{-1}$) & 401 & 70 & 670 & 418\\
Carrier conc. ($\times 10^{19}$cm$^{-3}$) & 2.38 & 7.5 & 1.6 & 2.95\\
Lattice Constant (a=b) (\AA) & 4.37 & 4.35 & 4.34 & 4.39\\
$\alpha$  & -0.19 & -0.23 & -0.24 & -0.24\\
l$_\varphi$ (nm) & 213 & 149 & 108 & 156 \\

\hline
\hline
\end{tabular}
\caption{Summary of structural and transport properties of  Bi$_2$Te$_3$ samples discussed in the manuscript.}
\label{table}
\end{table}

\begin{figure} [tbh]
    \centering
    \includegraphics[width=8cm]{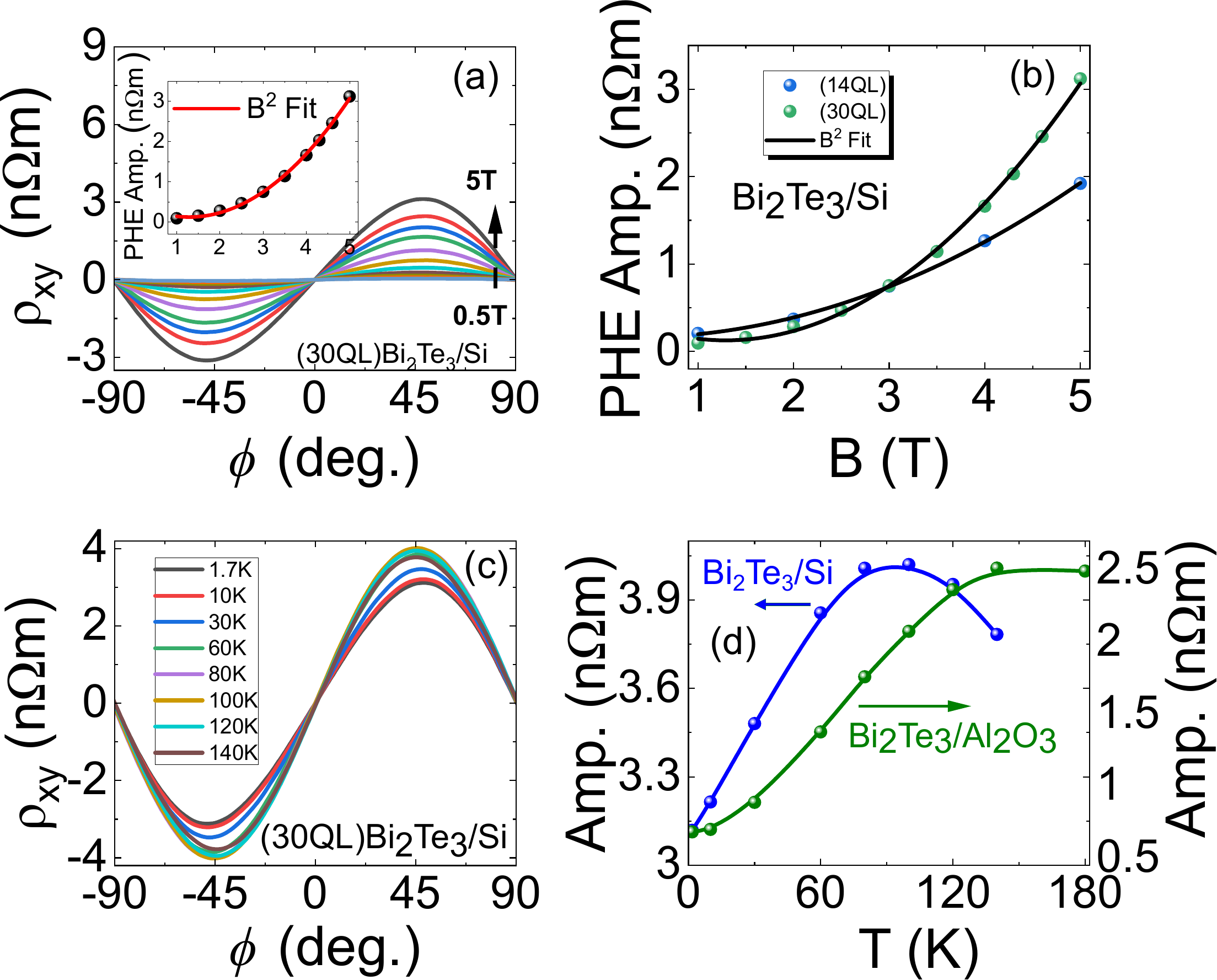}
    \caption{(a) $\rho_{xy}$ vs $\phi$ for BSi30, inset shows PHE amplitude vs field with B$^2$ fit.  (b) PHE amplitude vs B for BSi14 (blue) and BSi30 (green) with B$^2$ fit (black) (c) $\rho_{xy}$ vs $\phi$ for BSi30 at different temperatures  (d) PHE amplitude vs temperature for BSi30 and BA30. The down-turn in PHE amplitude in  BSi30 at T $>$ 100~K appear due to increasing conduction in Si at higher temperatures.}
    \label{30QL}
\end{figure}

Since it is important to understand if the origin of PHE stems from the  surface or the bulk, we examine  PHE in thicker samples (thickness = 30 QL) of Bi$_2$Te$_3$ maintaining exactly the same growth conditions as the respective  thinner samples. Fig.~\ref{30QL} (a) shows the transverse resistivity as a function of magnetic field, which exhibits a conventional behavior: PHE amplitude varies as B$^2$ and is sinusoidal with respect to $\phi$.  We compare the relative PHE amplitudes of the thicker and thinner film devices (fig.\ref{30QL} (b)). It is intriguing that the PHE amplitude of BSi30 sample is higher than the BSi14 at B = 5 T. However, the  magnitude of PHE amplitude in BA30 drops relative to BA14 (fig.~\ref{30QL}~(d)). Relative to the respective thinner samples, the mobility in BSi30 enhances $\approx$ 6 times as compared to a marginal enhancement ($\approx$ 1.7 times) in BA30; carrier concentrations of the samples do not vary significantly (refer table I). This suggests that bulk-disorder is the primary reason for the above contrasting trends in the magnitude of PHE amplitude for the two sets of samples (BA14-BA30 \& BSi14-BSi30). In BSi30, with a considerable drop in bulk-disorder, PHE signal increases due to larger bulk contribution. While in BA30, increase in thickness primarily reduces the surface contribution to the overall PHE amplitude explaining the drop in PHE amplitude relative to BA14. However, interestingly in both BSi30 and BA30, we observe an increase in PHE amplitude with increase in temperature  (fig.~\ref{30QL} (c)-(d)). This contrasting trend of PHE amplitude with temperature for thicker films compared to the thinner films corroborates our inference that here in thicker films bulk states predominantly contribute to the PHE signal.

\par We now discuss all plausible mechanisms that lead to PHE in Bi$_2$Te$_3$. Taskin et. al., \cite{Taskin2017} propose anisotropic lifting of topological protection of the surface states causing PHE. However, in our case it is evident that the  carrier concentrations of the samples do not correspond to the energy of the topologically protected surface state regime, where this theory could be applied. The picture of chiral anomaly is ruled out since chirality is not  a well defined quantity in 3D topological insulators. Nandy et. al.,\cite{Nandy2018} show that a 3D-TI with negative magnetoresistance, can exhibit PHE without the need to invoke chiral anomaly. This occurs purely because of the Berry curvature induced anomalous magneto-transport in the bulk limit.  In our samples, correlation between weak anti-localization and planar Hall effect seems to be less since WAL exists only at low temperatures whereas PHE persists up to relatively much higher temperatures.  S-H Zheng et. al., \cite{PhysRevB.101.041408} propose the existence of planar Hall voltage arising  due to non-linear terms in momentum due to spin-orbit coupling in the TI. By this theory, the emergence of PHE in 3D-TI seems to be of non-topological origin and arises purely due to the tilt of the Dirac cone in the presence of an in-plane magnetic field. From our experiments it is clear that PHE arises despite not being in the surface state regime of the band structure. We do not rule out the possibility that  PHE is  sensitive to the Fermi level position, however this requires a rigorous calculation of the PHE specific to the band structure of Bi$_2$Te$_3$. We believe that the PHE in  Bi$_2$Te$_3$ stems from the dominant bulk contribution, as evident from the observation of increase in PHE signal with in temperature in our thicker devices. While disentangling the bulk and the surface states has always been a challenge experimentally,  emergence of PHE in bulk dominated TI establishes that the effect does not necessarily require surface state transport. This calls for a theoretical calculation that addresses PHE specific to the band structure of Bi$_2$Te$_3$.  Our experiments pave the way for further theoretical investigation of anisotropies  leading to PHE in TIs.

In conclusion, our experiment  demonstrates a comprehensive study of PHE in 3D-TI Bi$_2$Te$_3$.  The PHE amplitude  varies as  square of the magnetic field and agrees well with the conventional PHE equations. Interestingly, the PHE amplitude enhances in BSi30 as compared to the BSi14. Furthermore, in contrast to the 14 QL samples, PHE amplitude in the 30 QL samples increases with temperature. While the origin of PHE in TIs is attributed to arise because of topologically protected surface states, our experiment provides a different perspective. We infer that the PHE has a significant contribution from bulk conduction as well, paving the way for further theoretical modelling of PHE in TIs that may assist in the future development of quantum technologies.
\vspace{-0.5cm}
\section*{Supplementary Material} Contains data of additional devices, fitting routine and description of calculations used in the manuscript.
\vspace{0.25cm}\\
\textit{Acknowledgments:} DS thanks Jagadeesh S. Moodera and Abhiram Soori for illuminating discussions.  All authors acknowledge extra-mural funding from TIFR-Hyderabad, Department of Atomic Energy, Government of India, under Project Identification No. RTI 4007,  SERB Sanction No. ECR/2015/000199 and CRG/2019/003810.
\vspace{0.25cm}\\
\textit{Data Availability}: The data that support the findings of this study are available from the corresponding author upon request.

\end{document}